\documentclass[11pt,preprint]{aastex61}

\usepackage{natbib}
\usepackage {float}
\usepackage{graphicx}
\usepackage{amssymb,amsmath,mathtools}

\received{}
\accepted{}
\shorttitle{DR2}
\shortauthors{Riess et al.}

\newcommand{\xddots}{%
  \raise 5pt \hbox {.}
  \mkern 6mu
  \raise 1pt \hbox {.}
  \mkern 6mu
  \raise -3pt \hbox {.}
}

\newcommand{\bq}{\begin{equation}} 
\newcommand{\eq}{\end{equation}}

\newcommand{\beq}{\begin{equation}}
\newcommand{\eeq}{\end{equation}}
\newcommand{\beqa}{\begin{eqnarray}}
\newcommand{\eeqa}{\end{eqnarray}}

\long\def\check#1{}
\long\def\hide#1{}

\begin{document} 

\title{Seven Problems with the Claims Related to the Hubble Tension in arXiv:1810.02595}

\author{Adam G.~Riess}
\affiliation{Space Telescope Science Institute, 3700 San Martin Drive, Baltimore, MD }
\affiliation{Department of Physics and Astronomy, Johns Hopkins University, Baltimore, MD}

\author{Stefano Casertano}
\affiliation{Space Telescope Science Institute, 3700 San Martin Drive, Baltimore, MD}
\affiliation{Department of Physics and Astronomy, Johns Hopkins University, Baltimore}

\author{D'Arcy Kenworthy}
\affiliation{Department of Physics and Astronomy, Johns Hopkins University, Baltimore, MD}

\author{Dan Scolnic}
\affiliation{Kavli Institute for Cosmological Physics, University of Chicago, Chicago, IL}
\affiliation{University of Chicago}

\author{Lucas Macri}
\affiliation{Texas A\&M University, Department of Physics and Astronomy, College Station, TX}

\begin{abstract} 

\cite{Shanks:2018} make two claims that they argue bring the local measurement and early Universe prediction of $H_0$ into agreement: A) they claim that Gaia DR2 parallax measurements show the geometric calibration of the Cepheid distance scale used to measure $H_0$ to be grossly in error and B) that we live near the middle of an enormous void, further biasing the local measurement of the Hubble constant.  We show that the first claim is caused by five erroneous uses of the data: in decreasing order of importance: 1) the use of a distance indicator, main sequence fitting of cluster stars, which is {\it unrelated} to the present calibration of Cepheids and therefore has no bearing on current measurements of $H_0$; 2) the use of Gaia data for Cepheids that fully saturate the detector, producing  unreliable parallaxes; 3) the use of a fixed parallax offset which is known to depend on source magnitude and color but which is derived for sources with extremely different colors and magnitudes; 4) ignoring the uncertainty in this offset; and 5) ignoring the other geometric sources of Cepheid calibration, the distance of the LMC from detached eclipsing binaries and the masers in NGC 4258, which are independent of Milky Way parallaxes.  Just resolving the first two of these issues by not using unrelated or saturated data leads to no inconsistency between Gaia parallaxes and the current Cepheid distance scale.  

The second claim can be refuted 6) because of the increase in $\chi^2$ that the alleged void would entail in SN measurements in the Hubble flow, and 7) because it would represent a 6 $\sigma$ fluctuation of cosmic variance between the local and globally measured expansion, requiring us to live in an exceedingly special location.

\end{abstract} 

\keywords{astrometry: parallaxes --- cosmology: distance scale --- cosmology:
observations --- stars: variables: Cepheids --- supernovae: general}

\section{Introduction} 

The locally measured value of the Hubble constant differs with 3.6 $\sigma$ confidence from its value predicted using the Planck CMB data to calibrate the $Lambda CDM$ model of the Universe \citep{Planck:2018,Riess:2018b,Birrer:2018}.  This difference is widely referred to as the ``Hubble Tension".  Recently, \cite{Shanks:2018} have made two claims that they argue bring the local measurement and early Universe prediction of $H_0$ into closer agreement.  They claim A) that Gaia DR2 parallaxes of Milky Way Cepheids imply that the geometric calibration of the local distance scale used to measure $H_0$ is grossly in error and B) that we live near the middle of an enormous void, further biasing the local measurement of the Hubble constant.   Here we present seven  problems in their analysis which render their conclusions invalid.

\section{Claim A: The Geometric Calibration of the Distance Ladder}

The first rung of the current Cepheid---SN Ia distance ladder is calibrated through 5 independent sources of geometric distances to Cepheids: 1) Keplerian motion of water masers in NGC 4258; 2) 8 Detached Eclipsing Binary systems in the LMC; 3) parallaxes of bright Milky Way Cepheids from \cite{benedict07} measured with the HST FGS; 4) parallaxes of long-period Milky Way Cepheids from \cite{Riess:2018a}, measured via spatial scanning of WFC3 on HST; and 5) Milky Way parallaxes \citep{Riess:2018b} measured in Gaia DR2.  The value of the Hubble constant from each geometric source and its contribution to the error in $H_0$ are: 72.3 $\pm$ 1.9, 72.0 $\pm 1.8$, 76.2 $\pm 1.7$, 75.7 $\pm 2.5$ and 73.7 $\pm 2.4$, respectively, with the errors representing only the respective uncertainty of the Cepheid calibration (and of the local distance scale thus derived).  These values are consistent with the combined value of 73.5 given in \cite{Riess:2018b}; each supports the tension with the Planck and $\Lambda CDM$ derived prediction of 67.4 $\pm 0.5$.   The geometric calibration of Cepheid luminosities presented in \cite{Riess:2016} were based on only the first three of these, and considerations in \cite{Shanks:2018} do not apply to the first two of these three sources.

\cite{Shanks:2018} compare a set of 3 types of previous Cepheid distance measurements to the Gaia DR2 parallaxes after applying the QSO-derived parallax offset and claim a large inconsistency, implying an error in the Cepheid luminosity and thus in the local distance scale.  They further infer that the determination of $H_0$ is in error by a similar amount.  Their analysis is vitiated by the five serious problems discussed below, which make their conclusions invalid.

\begin{itemize}

\item[1)] About half of the Cepheids they consider (14 of 29) are members of open clusters, and their previous distances are based on {\it main sequence fitting of clusters} (MSFC), published by \cite{Laney:1993}, and calibrated on the basis of Allen's Astrophysical Quantities (1973).  {\it These distances have never been used in modern $ H_0 $ determination, and their values are thus irrelevant in the context of distance scale.}  All Cepheid distances used in the determination of $H_0$ are geometric.  The mean difference of $ 0.34 \pm 0.08 $ only pertains to the MSFC method.  Applying this difference to modern $H_0$ determinations is a {\it fundamental logical fallacy}; the previous MSFC distances could have been found in error by any arbitrary amount, with {\it no impact whatsoever} on the distance scale.

\item[2)] About a quarter of the sources (8 of 29) are Milky Way Cepheids with parallaxes measured by HST FGS \citep{benedict07}.  All of these have a mean magnitude $G<6$ mag during some or all of their light curves; therefore they saturate in Gaia astrometric observations.  Saturation is known to produce spurious parallaxes due to as yet imperfect modeling of the PSF; the Gaia Team strongly recommends parallaxes for saturated objects in DR2 not be used, calling them ``unreliable" \citep{Lindegren:2018}.

Simply removing the above unrelated or saturated data from the \cite{Shanks:2018} analysis leads to no inconsistency between Gaia parallaxes and the current Cepheid distance scale from the calculations given in \cite{Shanks:2018}.

\item [3)] The remaining sources (7 of 29) are Milky Way Cepheids with HST spatial scan parallaxes.  This is the only valid comparison set, as they were in fact used in the determination of $H_0$ \cite{Riess:2018b} and the Gaia DR2 measurements are not flagged by saturation.  

   Here further explanation is required as Gaia DR2 parallaxes are the only one of five independent geometric anchors of the Cepheid calibration \citep{Riess:2018b} that require calibration for a specific application before use.  The Gaia Team \citep{Lindegren:2018} (L18) explain that to obtain a true parallax it is necessary to add a parallax ``zeropoint'' offset term whose appropriate value depends on source magnitude, color and position on the sky.  According to L18, “the actual offset applicable for a given combination of magnitude, color, and position may be different by several tens of $\mu$as.”  The main reason for this offset is an unexpectedly large variation in the Gaia basic angle.  L18 shows that the median parallax offset of blue QSOs with $14 < G < 21$ mag (their Figure 7) is $ -29 \,\mu$as, with a potential increase in absolute value, to $\sim -50 \,\mu$as, for brighter and redder QSOs. Because Milky Way Cepheids are $\sim$ 5 magnitudes brighter than even the brightest QSOs, the QSO sample simply cannot be used to determine the parallax offset for Milky Way Cepheids. 

For this subset, \cite{Shanks:2018} find a difference in distance scale of 0.21 $\pm$ 0.18 mag between the Cepheids and the Gaia parallaxes {\it when assuming the parallax offset of $ -29 \,\mu$as determined from the bulk of the QSOs}.  However, as stated above, even within the QSO sample the parallax offset appears to vary by several tens of $\mu$as; with an offset value of $\sim -50 \,\mu$as, more appropriate for brighter and redder QSOs, there is {\it no difference} in distance scale between the 7 Cepheids and their Gaia parallaxes.  \cite{Riess:2018b} used a set of 46 Milky Way Cepheids with accurate HST photometry in the same system used for extragalactic Cepheids to simultaneously solve for an additive term (the parallax offset) and a multiplicative term (an error in the distance scale); they found a parallax zeropoint offset of $ -46 \pm 13$ $\mu$as and no offset in distance scale vs. all other anchors.  \cite{Zinn:2018} find an offset of $ -53 \pm$ 2.6 $\mu$as on the parallax zeropoint offset calculated from 3475 red giants with Kepler-based asteroseismic estimates of radii and parallaxes, stars with magntudes and color much closer to Milky Way Cepheids than QSOs.  As a further comparison, we selected a sample of 600 Milky Way Cepheids identified in the Gaia DR2 Cepheid catalog, including only objects with full photometric and parallax information and classified as fundamental-mode,  $ \delta $ Cep-type  Cepheids.  For these Cepheids, a comparison between photometric and trigonometric parallaxes (Figure 1) clearly shows that the parallax offset is consistent with the QSO offset for $ G \sim 16 $, but increases (in absolute value) across the brightness range not sampled by QSOs, to values approaching $ - 80 $ $\mu$as close to the saturation limit.   

\item [4)] \cite{Shanks:2018} assume no uncertainty in the parallax offset despite the fact that L18 show it to vary by several tens of $\mu$as depending on source magnitude and color. Simply including an uncertainty of this size would demonstrate the lack of significance for an offset since the difference between the QSO and Cepheid or Red Giant-derived value {\it is this size}. No explanation is given by \cite{Shanks:2018} for claiming no uncertainty to this weakly constrained parameter.

\item [5)] The geometric calibration of Cepheid luminosities in R16 is based on two other sources, detached eclipsing binaries in the LMC and water masers in NGC 4258, both of which are independent of any Milky Way parallaxes.  The claims by \cite{Shanks:2018} would not impact and would not warrant a revision of $H_0$.

\end{itemize}

To summarize, simply considering the first two issues, i.e., excluding the immaterial comparison with MSFC distances and Cepheids that saturate in Gaia, makes the supposed discrepancy of no significance ($ 0.21 \pm 0.18 $ mag).  What remains directly results from the application of an inappropriate zeropoint offset, as documented by the Gaia Team (L18), or any allowance for its substantial uncertainty.

\section{Claim B: The Local Hole}

\cite{Shanks:2018} cites galaxy studies from \cite{Whitbourn:2014} that show that over the $\sim$20\% of the sky they studied space is denser at $z>0.05$ than at $z<0.05$.  They then assume that this feature persists over the 80\% of the sky not similarly studied, i.e, they {\it assume a spherical structure} with the Milky Way located in a ``local hole".   Its is important to stress that this model is dominated by an assumption, that 80\% of the local Universe can be inferred from seeing 20\% and guessing at a pattern, i.e., spherical symmetry centered on or near the Milky Way.

\begin {itemize}

\item [6)]  \cite{Shanks:2018}  use the Pantheon sample \citep{Scolnic:2018} of 1048 SNe to see whether this feature is present in the Hubble diagram of SNe.  They find a total $\chi^2$ of 1037.7 assuming no hole and 1049.3 with the hole included; this hole would lower the Hubble constant from \cite{Riess:2016} by 1.8\%.  The conventional (i.e., frequentist) interpretation of a comparison where a model feature {\it increases} the total $\chi^2$ by 11.5 is that it is excluded with 3.4 $\sigma$ confidence (99.99\%).  Kenworthy et al (2018) {in prep} show from the combination of Pantheon, Foundation \citep{Foley:2018,Jones:2018} and CSP III \citep{Krisciunas:2017} SNe samples that this local hole model is excluded at 4.5 $\sigma$.  Because the local hole model is 80\% assumption and strongly disfavored by the Hubble diagram of SNe, a change in $H_0$ is not warranted.

\item [7)] The local hole claim by \cite{Shanks:2018} would also imply that we live in a 6 $\sigma$ fluctuation in the cosmic variance of Hubble constant measurements, on the basis of studies by \cite{Wu:2017} and \cite{Odderskov:2017} who find the cosmic variance of the local measurement of $H_0$ to be 0.3\%.  It would also require us to live in an exceedingly special place.

\end {itemize}

\begin{figure}[ht]
\vspace*{150mm}
\figurenum{1}
\includegraphics{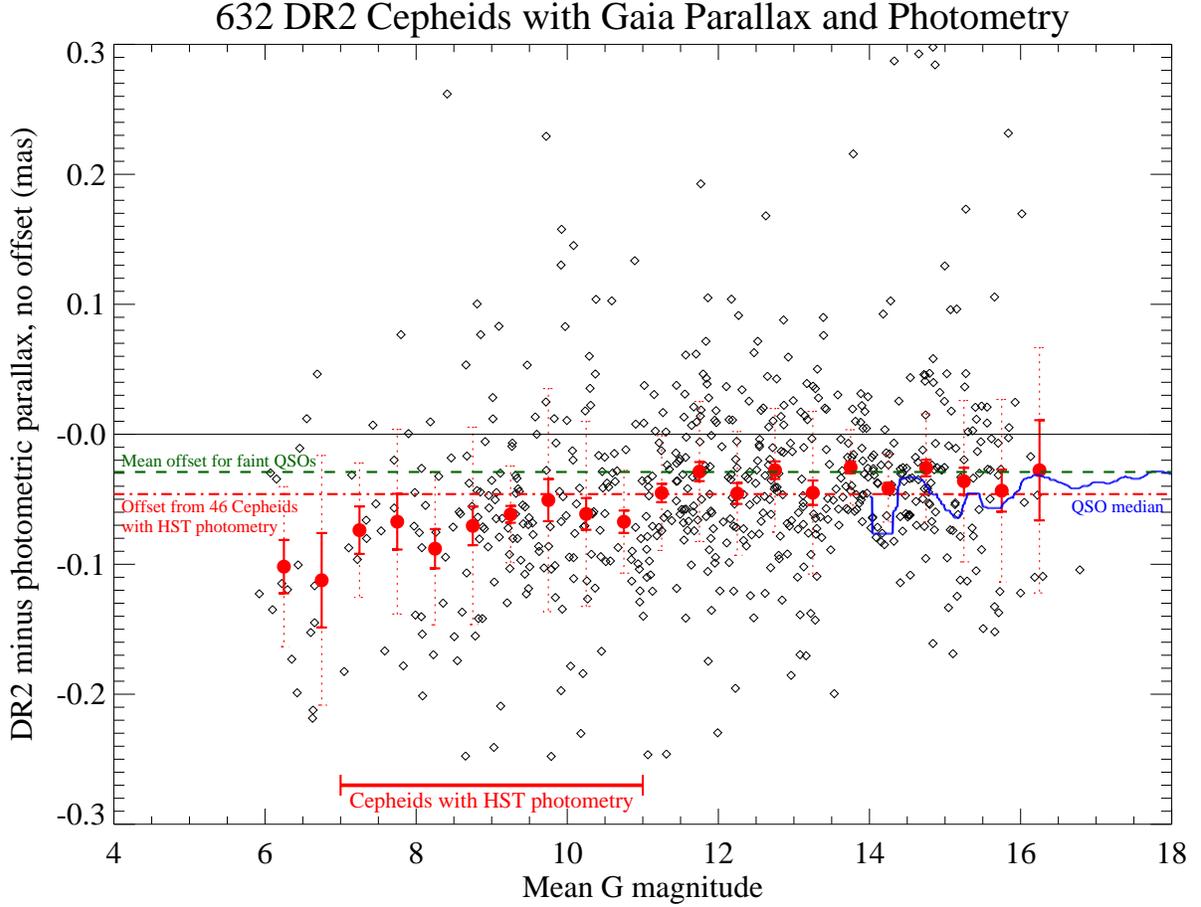}
\caption{\label{fg:outlier}  Astrometric and photometric parallax comparison for the 632 Milky Way Cepheids that Gaia classifies as Fundamental mode, Delta-Cepheii type, not in the LMC or SMC or those in the secondary sequence $\sim$5 mag fainter.  The Wesenheit magnitudes \citep{madore82} are calculated from Gaia G, Bp, and Rp magnitudes regressing with Cepheids in common with HST system magnitudes \citep{Riess:2018b} to derive transformations between the photometric systems.  The absolute photometric parallaxes are on the \cite{Riess:2016} Cepheid distance scale.
The Gaia DR2 parallaxes do not include the parallax zeropoint offset which must be added to calibrate Gaia parallaxes \citep{Lindegren:2018} and which is known to depend on source magnitude and color \citep{Lindegren:2018}.  As shown, the offset for 46 Milky Way Cepheids from \cite{Riess:2018b} is -46 $\pm 13$ $\mu$as (red line) and larger for the brightest Cepheids, consistent with the bright end of the QSO's (blue line), Red Giants \citep{Zinn:2018} and Cepheids on the HST system \citep{Riess:2018b} but clearly larger than the mean of $-29$ $\mu$as (green line), derived from the mean QSO which is bluer and much fainter ($G\sim19$) mag than Cepheids.  The Cepheids are binned for easy comparison.  \cite{Shanks:2018} has assumed the offset from the median QSO for the 7 HST Cepheids with mean $G=9$ mag \citep{Riess:2018a} which is in poor agreement with the 632 Cepheids and the bright end of the QSO's producing an offset in the calculated distance scale.}
\end{figure}

\bigskip

\bigskip

\acknowledgements

 \vfill
\eject

\clearpage
\bibliographystyle{aasjournal} %
\bibliography{bibdesk}
\clearpage

\end{document}